\documentclass[preprint,prd,showpacs,superscriptaddress]{revtex4}

\usepackage{graphicx}

\begin{document}

\title{\bf Quantum Stress Tensor Fluctuations of a Conformal Field
and Inflationary Cosmology}

\begin{flushright}
CECS-PHY-10/6, UFIFT-QG-10-02
\end{flushright}

\author{L.H. Ford}
\email{ford@cosmos.phy.tufts.edu}
\affiliation{Institute of Cosmology,
Department of Physics and Astronomy \\
Tufts University, Medford, MA 02155 USA}
\author{S.P. Miao}
\email{smiao@cecs.cl}
\affiliation{Centro de Estudios Cient\'ificos (CECS),
Casilla 1469, Valdivia, Chile}
\author{Kin-Wang Ng}
\email{nkw@phys.sinica.edu.tw}
\affiliation{Institute of Physics,
Academia Sinica, Nankang, Taipei 11529 Taiwan}
\author{R.P. Woodard}
\email{woodard@phys.ufl.edu}
\affiliation{Department of Physics, University of Florida,
Gainesville, FL 32611 USA}
\author{Chun-Hsien Wu}
\email{chunwu@phys.sinica.edu.tw}
\affiliation{Institute of Physics,
Academia Sinica, Nankang, Taipei 11529 Taiwan}
\affiliation{Department of Physics, Soochow University,
70 Linhsi Road, Shihlin, Taipei 111 Taiwan}

\begin{abstract}
We discuss the additional perturbation introduced during inflation
by quantum stress tensor fluctuations of a conformally invariant field
such as the photon. We consider both a kinematical model, which deals
only with the expansion fluctuations of geodesics, and a dynamical
model which treats the coupling of the stress tensor fluctuations to a
scalar inflaton. In neither model do we find any growth at late times,
in accordance with a theorem due to Weinberg. What we find instead is
a correction which becomes larger the earlier one starts inflation.
This correction is non-Gaussian and highly scale dependent, so the
absence of such effects from the observed power spectra may imply a
constraint on the total duration of inflation. We discuss different
views about the validity of perturbation theory at very early times
during which currently observable modes are transplanckian.
\end{abstract}


\maketitle

\baselineskip=24pt 

\section{Introduction}

The inflationary paradigm has been remarkably successful in
predicting observed features of the cosmic microwave background~(CMB) 
radiation and the large scale structure of the Universe. If inflation
is driven by a nearly free, massless quantum field, then a generic
prediction is a spectrum of primordial fluctuations which is Gaussian
and almost scale-invariant~\cite{MC81,GP82,Hawking82,Starobinsky82,BST83}.
For a recent review, see for example~\cite{Mukhanov}. The best
test of these predictions comes from CMB observations by the WMAP
satellite, which has found a spectrum of temperature fluctuations
consistent with Gaussian, nearly scale-invariant primordial 
fluctuations~\cite{WMAP}. 

However, in addition to the dominant effect coming from tree order
fluctuations of the scalar inflaton and of the graviton, there
should also be some effects from loop corrections of these fields
with themselves and with other fields.
The latter will be the topic of this paper, particularly a one loop
effect which can be interpreted in terms of quantum stress tensor
fluctuations.  The fluctuations of
quantum stress tensors and their physical effects have been discussed
by several authors in recent 
years~\cite{WF01,Borgman,Stochastic,FW04,TF06}. For a recent review
with further references, see Ref.~\cite{FW07}. Quantum stress tensor 
fluctuations necessarily have a skewed, highly non-Gaussian,
probability distribution, although the explicit form of this
distribution has only been found in two-dimensional spacetime 
models~\cite{FFR10}.

Ref.~\cite{WNF07} studied the possible contributions of quantum stress tensor
fluctuations of a conformally invariant field to primordial density 
perturbations in inflationary
models. It was found that these contributions can be proportional
to a power of the scale factor change during inflation, and hence
potentially large enough to observe. Because they are associated with
a non-scale invariant and non-Gaussian contribution, they can at best be
a sub-dominant part of the  primordial density perturbations. This
fact was used in Ref.~\cite{WNF07} to infer upper bounds on the
duration of inflation. These bounds are compatible with adequate
inflation to solve the horizon and flatness problems, but raise the
possibility that the total duration of inflation might be observable.
This possibility goes against a commonly held view that inflation
erases the memory of anything which occurred previously, and hence
increasing its duration beyond the minimum needed to solve the 
horizon and flatness problems can produce no observable effect.
However, contrary indications to this view had previously been
published in the form of arguments that inflation cannot be eternal
to the past\cite{BV94,BGV03}, although these arguments were based on
general considerations which do not make  specific predictions
of observable effects. Winitzki~\cite{W10} has recently suggested a 
model in which inflaton field fluctuations can produce violations of the
null energy condition and possible effects of the total inflationary
expansion.

The purpose of the present paper is to re-examine and improve
the analysis in Ref.~\cite{WNF07}. In Sect.~\ref{sec:kin}, we
discuss a kinematic model which makes no explicit reference to
the inflaton field, but examines the gravitational effects of the
stress tensor fluctuation upon timelike geodesics. In 
Sect.~\ref{sec:dyn}, we give a detailed treatment of a dynamical
model in which the stress tensor fluctuations alter the dynamics of a
scalar inflaton field. In both models, a correction to the power 
spectrum of density fluctuations is computed. Section~\ref{sec:density} 
discusses the implications of our results and some associated conceptual 
issues, especially the role of transplanckian modes. Our analysis is 
summarized in Sect.~\ref{sec:final}.

Before concluding this section we should mention some conventions. A
hat is used to denote the spatial Fourier transform of any field
$A(t,{\bf x})$
\begin{equation}
\hat{A}(t,{\bf k}) \equiv \frac1{(2 \pi)^3} \int \!\! d^3x \,
e^{i {\bf k} \cdot {\bf x}} A(t,{\bf x}) \; .
\end{equation}
We represent the {\it power spectrum} of $A$ by the symbol 
${\cal P}_A(k,t)$, which is defined as follows from the correlator of 
two $\hat{A}$ fields:
\begin{equation}
\langle \hat{A}(t,{\bf k}) \hat{A}(t,{\bf k}') \rangle \equiv
{\cal P}_A(k,t) \times \frac{\delta({\bf k} \!+\! {\bf k}')}{4 \pi k^3} 
\; . \label{power}
\end{equation}
In (the usual) cases for which $\hat{A}(t,{\bf k})$ is time independent
we drop time from the argument list of the power spectrum, as in
${\cal P}_A(k)$. We consider the loop counting parameter of quantum 
gravity to be $\kappa^2 \equiv 16 \pi G$. Our curvature tensors follow the
Landau-Lifshitz spacelike convention, which is also the
Misner-Thorne-Wheeler (+++) convention,

\begin{equation}
R^{\rho}_{~\sigma\mu\nu} \equiv \partial_{\mu} \Gamma^{\rho}_{~\nu\sigma}
- \partial_{\nu} \Gamma^{\rho}_{~\mu\sigma} + \Gamma^{\rho}_{\mu \alpha}
\Gamma^{\alpha}_{~\nu\sigma} - \Gamma^{\rho}_{~\nu\alpha}
\Gamma^{\alpha}_{~\mu\sigma} \qquad {\rm and} \qquad R_{\mu\nu} \equiv
R^{\rho}_{~ \mu\rho\nu} \; .
\end{equation}
A very important point for understanding our analysis and results is
that we normalize the FRW scale factor to unity at the end of inflation,
rather than at the current time. Also note that we use the subscript ``0''
sometimes to signify ``background'' and sometimes to denote that the
subscripted quantity is evaluated at the beginning of inflation. So $t_0$
is the time at which inflation begins, rather than the current time as in 
much of the literature on cosmology. We indicate the current time by the
subscript ``now'', so the wave number $k = 2\pi/\lambda$ is measured in
units of the comoving distance at the end of inflation, and it can
be expressed in terms of the current wave number $k_{\rm now} = 
2\pi/\lambda_{\rm now}$ through the relation $k = a_{\rm now} k_{\rm now}$.

\section{The Kinematic Model Revisited}
\label{sec:kin}

\begin{figure}
\begin{center}
\includegraphics[width=3.5cm,height=1.6cm]{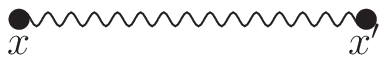}
\hspace{.5cm}
\includegraphics[width=3.5cm,height=1.6cm]{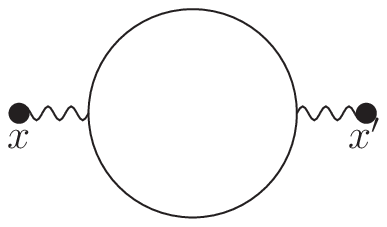}
\hspace{.5cm}
\includegraphics[width=3.5cm,height=1.6cm]{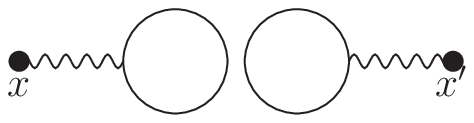}
\end{center}
\caption{Various contributions to the power spectrum of primordial
perturbations. Wavy lines stand for graviton-inflaton fields and solid
lines denote conformal fields. The leftmost diagram represents the tree
order contribution which is usually reported. The center diagram gives
the one loop contribution from conformal matter which is the subject of
our analysis. The rightmost diagram represents the (unobserved) term
which is neglected by subtracting off the expectation value of the
conformal stress tensor.}
\label{diagrams}
\end{figure}

Here we will review and modify a model first presented in Ref.~\cite{WNF07}. 
The point is to give a simple computation of the extra part of the power
spectrum of energy density fluctuations due to a conformally invariant
quantum field. (See Fig.~\ref{diagrams} for the relation between our
contribution and the usual tree order result.) A rigorous derivation 
involves solving the coupled, linearized inflaton-graviton equations with 
the conformal stress tensor as a source. We will do that in 
Sect.~\ref{sec:dyn}. Here we avoid any mention of the inflaton field
and we require only the background metric
\begin{equation}
ds^2= -dt^2 +a^2(t) \; \delta_{ij} dx^i dx^j 
    = a^2(\eta)\,(-d\eta^2  + \delta_{ij} dx^i dx^j )\,,
\label{eq:RW}
\end{equation}
where $t$ is the comoving time, $\eta$ is the conformal time, 
and $a$ is the scale factor. 

What we do instead is to assume that the stress energy consists of a perfect
fluid with energy density $\rho(t,{\bf x})$, pressure $p(t,{\bf x}) = w
\rho(t,{\bf x})$ (with constant equation of state $w$) and 4-velocity
$u^{\mu}(t,{\bf x})$, in co-moving coordinates such that $u^{\mu}
\partial_{\mu} = {\partial}/{\partial t}$. Then we use energy conservation,
\begin{equation}
\dot{\rho} + (\rho + p) \theta = 0 \; , \label{eq:conservation}
\end{equation}
with $\dot{\rho} \equiv \partial \rho/\partial t$,
to infer the perturbed energy density $\delta \rho(t,{\bf x})$ by
perturbing the expansion $\theta(t,{\bf x}) \equiv u^{\mu}_{~ ;\mu}$
\begin{equation}
\frac{\partial}{\partial t} 
\left( \frac{\delta \rho(t,{\bf x})}{\rho_0(t)} \right)
= -(1 + w) \delta \theta(t,{\bf x}) \; .
\end{equation}
The key to simplifying the computation is deriving the perturbed expansion
$\delta \theta(t,{\bf x})$ from the Raychaudhuri equation
\begin{equation}
u^{\mu} \partial_{\mu} \theta = -R_{\mu\nu} u^{\mu} u^{\nu} -
\frac13 \theta^2 - \sigma_{\mu\nu} \sigma^{\mu\nu} + \omega^{\mu\nu}
\omega^{\mu\nu} + (u^{\mu}_{~ ;\nu} u^{\nu})_{;\mu} \; . \label{Ray}
\end{equation}
We shall drop the shear $\sigma^{\mu\nu}$, the vorticity $\omega^{\mu\nu}$,
and the acceleration $(u^{\mu}_{~ ;\nu} u^{\nu})_{;\mu}$, at which point one 
can obtain $\delta \theta(t,{\bf x})$ from $\delta R_{\mu\nu}(t,{\bf x})
= \frac12 \kappa^2 T_{\mu\nu}^{\rm conf}(t,{\bf x})$, for the particular 
part of the total perturbation that concerns us. (See
Fig.~\ref{diagrams}.) Here $T_{\mu\nu}^{\rm conf}$ denotes the stress
tensor of the conformal field.

This makes for a wonderfully simple analysis in which we need never
consider the perturbed inflaton field or components of the perturbed
metric. Unfortunately, it isn't correct, as we will see in Sect.~\ref{sec:dyn}.
Ignoring $\sigma^{\mu\nu}$ and $\omega^{\mu\nu}$ is valid at linearized
order for single-scalar inflation, but the acceleration term contributes
at linearized order and that spoils the simple relation between $\delta
\theta(t,{\bf x})$ and the conformal stress tensor. So the result we
shall derive in this section is off by an important factor of $(k/H)^4$,
but it does depend correctly on the initial time.

Conformal invariance allows the stress tensor correlation function
\begin{equation}
C_{\mu\nu\alpha\beta}(x,x') = 
\langle  T_{\mu\nu}(x)\, T_{\alpha\beta}(x') \rangle
- \langle  T_{\mu\nu}(x) \rangle \langle T_{\alpha\beta}(x') \rangle
\,,
\label{eq:corr-fnt}
\end{equation} 
to be written in terms of the flat space stress tensor correlation 
function.
\begin{equation}
C_{\mu\nu\alpha\beta}^{RW}(x,x') = a^{-2}(\eta)\,a^{-2}(\eta')\,
C_{\mu\nu\alpha\beta}^{flat}(x,x')\, .
\end{equation}  
Here the components of $C_{\mu\nu\alpha\beta}^{RW}(x,x')$ are
understood to be in the second set of coordinates in Eq.~(\ref{eq:RW}).
Although the conformal anomaly term in $\langle  T_{\mu\nu}(x)
\rangle$ breaks conformal symmetry, this term cancels out of the
correlation function, Eq.~(\ref{eq:corr-fnt}).

The flat spacetime energy density correlation
function of the conformal field is ${\cal E}(\Delta\eta,r)$, where
 $\Delta\eta = \eta -\eta'$ and $r = |\mathbf{x} - \mathbf{x'}|$.
The expansion correlation function can be
expressed in terms of  ${\cal E}(\Delta \eta,r)$ as
\begin{equation}
\langle \delta\theta(\eta_1,{\bf x})\, \delta\theta(\eta_2,{\bf x'}) \rangle =
\frac14 \kappa^4 \; a^{-2}(\eta_1)\,a^{-2}(\eta_2)\, \int_{\eta_{0}}^{\eta_{1}}
\frac{d\eta}{a(\eta)}
\int_{\eta_{0}}^{\eta_{2}}
\frac{d\eta'}{a(\eta')}\; {\cal E}(\Delta\eta,r) \,.
   \label{eq:theta_corr}
\end{equation}
For the case of the electromagnetic field,
\begin{equation}
{\cal E}_{em}(\Delta\eta,r)= {\rm Re}\left\{
\frac{(\Delta\eta^{2}+3r^{2})(r^{2}+3\Delta\eta^{2})}
{\pi^{4}[r^{2} - (\Delta \eta \!+\! i\varepsilon)^{2}]^{6}}\right\} \,,
\label{eq:em_corr}
\end{equation}
and the expression for the conformal scalar field case is identical except
for an additional factor of $1/12$. (Note that this expression corrects
an error in Eq.~(39) of Ref.~\cite{WNF07}.) This expression is ultraviolet
finite in spite of being one loop~\cite{FW04}.

After reheating, the expansion fluctuations cause differential redshifting
and consequent density fluctuations, in accordance with the
conservation law Eq.~(\ref{eq:conservation}) for a perfect fluid. The fluid 
flow approach to density perturbations has been discussed by several 
authors~\cite{Hawking66,Olson,LS90,LL93}. Let $\delta_\rho = \delta
\rho/\rho$ be the fractional density fluctuation at conformal time 
$\eta=\eta_s$, the last scattering surface. Its spatial correlation function
is given in terms of the expansion correlation function  by
\begin{equation}
\langle \delta_\rho(\eta_s,{\bf x})\,  \delta_\rho(\eta_s,{\bf x'})  \rangle=
 (1+w)^{2}\int_{\eta_r}^{\eta_{s}}\frac{d\eta_1}{a(\eta_1)}
\int_{\eta_r}^{\eta_{s}} \frac{d\eta_2}{a(\eta_2)} \,
\langle \delta\theta(\eta_1,{\bf x})\, \delta\theta(\eta_2,{\bf x'}) \rangle\,.
\label{eq:del_rho}
\end{equation} 
Here reheating occurs at  $\eta=\eta_r$ and the equation of state after
reheating is $p = w\, \rho$.

Let 
\begin{equation}
F_0(r) = \langle \delta\theta(\eta_r,{\bf x})\, 
                       \delta\theta(\eta_r,{\bf x'}) \rangle
\label{eq:F0}
\end{equation}
be the variance of the expansion at the end of inflation. 
In many cases, the dominant contribution to the density fluctuations
arises from effects occurring during inflation. Then contributions to
the expansion correlation function coming from stress tensor
fluctuations after reheating may be neglected. However, we still 
need to account for the evolution of $\delta\theta$ after reheating.
If we ignore the effects of classical density perturbations and
pressure gradients, as well as the quantum stress tensor, then
$\delta\theta$ satisfies [See Eq.~(13) in Ref.~\cite{WNF07}.]
\begin{equation}
\frac{d\delta \theta}{dt}= -
\frac{2}{3}\theta_{0} \, \delta\theta\,,
\label{eq:theta1}
\end{equation}
where $\theta_0 = 3 \dot{a}/a$ is the unperturbed Robertson-Walker
expansion. The solution of this equation can be written as
\begin{equation}
\delta \theta(\eta) = \frac{\delta \theta(\eta_r)}{a^2(\eta)} 
\,, \qquad \eta \geq \eta_r \,,
\end{equation}
where we have set the scale factor at reheating to unity, $a(\eta_r)=1$. 
In the approximation where we consider only stress tensor fluctuations
during inflation, after reheating we have
\begin{equation}
\langle \delta\theta(\eta_1,{\bf x})\, \delta\theta(\eta_2,{\bf x'}) \rangle =
 a^{-2}(\eta_1)\,a^{-2}(\eta_2)\; F_0(r)\,.
\end{equation}
 
The density perturbation correlation function now becomes
\begin{equation}
\langle \delta_\rho(\eta_s,{\bf x})\,  \delta_\rho(\eta_s,{\bf x'})  \rangle
\approx (1+w)^{2} \,F_0(r)\, 
\left[\int_{\eta_r}^{\eta_{s}}\frac{d\eta_1}{a^3(\eta_1)} \right]^2  \,.
\end{equation}  
The time integral now depends only upon the form of the scale factor
between reheating and last scattering. 
We consider a model in which the inflation is described by a de Sitter
metric,
\begin{equation}
a(\eta) = -\frac{1}{H \eta}\,, \qquad \eta \leq \eta_r \,,
\label{eq:deSitscale}
\end{equation}
and the subsequent period is radiation dominated ($w=1/3$),
\begin{equation}
a(\eta) = H\eta +2 \,.
\end{equation}
Here $H$ is the Hubble parameter of de Sitter space, and both $a(\eta)$ and
$da/d\eta$ are continuous at $\eta = \eta_r = -1/H$. Then
\begin{equation}
\int_{\eta_r}^{\eta_{s}}\frac{d\eta_1}{a^3(\eta_1)} =
\int_{-1/H}^{\eta_{s}}\frac{d\eta_1}{(H\eta_1 +2)^3} \approx \frac{1}{2\,H}\,.
\end{equation}
The last step follows because $a(\eta_s) \gg 1$. Now we obtain
\begin{equation}
\langle \delta_\rho(\eta_s,{\bf x})\,  \delta_\rho(\eta_s,{\bf x'})
\rangle = \frac{4}{9H^2} \,F_0(r)\, .
\label{eq:del_rho3}
\end{equation}

We will next take spatial Fourier transforms and write, for example,
\begin{equation}
\hat{\cal {E}}_{em}(\Delta\eta,k) =  \frac{1}{(2\pi)^3}\int d^{3}x\, 
{\rm e}^{i\,\mathbf{k}\cdot\Delta\mathbf{x}} \, {\cal E}_{em}(\Delta\eta,r)\,.
\label{eq:fourier}
\end{equation}
The explicit form of $\hat{\cal E}_{em}(\Delta\eta,k)$ is
\begin{equation}
\hat{\cal E}_{em}(\Delta\eta,k) = 
-\frac{k^4 \, \sin(k\, \Delta \eta)}{960 \pi^5\, \Delta \eta}
= -\frac{k^5}{960 \pi^5}\; \int_0^1 du \, \cos(k u \, \Delta \eta) \,.
\label{eq:em_corr_k}
\end{equation}
The Fourier transform of $F_0(r)$ is
\begin{equation}
\hat{F_0}(k) = \frac14 \kappa^4 \;  \int_{\eta_{0}}^{\eta_{r}}
\frac{d\eta}{a(\eta)} \int_{\eta_{0}}^{\eta_{r}}
\frac{d\eta'}{a(\eta')}\; {\cal E}(\Delta\eta,k) \,.
\label{eq:F0_k}
\end{equation}

We next evaluate the integrals in this expression, using the second
form in Eq.~(\ref{eq:em_corr_k}), to find in the limit that
$k|\eta_0| \gg 1$, 
\begin{equation}
\hat{F_0}(k) \sim \frac{\kappa^4 k^4 H^2}{11520 \, \pi^4}\, 
\left( - |\eta_0|^3 + \frac{3}{\pi k}\, |\eta_0|^2 + \cdots \right)\,.
\label{eq:F0_asym}
\end{equation}
Note that the magnitude of $\hat{F_0}(k)$ grows as $|\eta_0| \rightarrow 
\infty$ for fixed $k$. The growth was found in Ref.~\cite{WNF07}, but 
there only the $|\eta_0|^2$ term appeared. The reason for this is that
in Ref.~\cite{WNF07}, the calculations were done in coordinate space
until the last step, where a term in $F_0(r) \propto 1/r^6$ was
Fourier transformed into a term in $\hat{F_0}(k) \propto k^3$. However, this
procedure is not sensitive to the possibility of delta-function terms
in $F_0(r)$. The leading term in Eq.~(\ref{eq:F0_asym}) arises
from just such a term, one proportional to 
$\nabla^4 \delta(\mathbf{x} - \mathbf{x'})$.

Let $P_{\delta_{\rho}}(k)$ denote the spatial Fourier transform of
$\langle \delta_\rho(\eta_s,{\bf x})\, \delta_\rho(\eta_s,{\bf x'})
\rangle$, which is related to the power spectrum by  
${\cal P}_{\delta_{\rho}}(k) = 4 \pi k^3\,P_{\delta_{\rho}}(k)$. 
From Eqs.~(\ref{eq:del_rho3}) and (\ref{eq:F0_asym}), we find
\begin{equation}
P_{\delta \rho}(k) \equiv \frac1{(2\pi)^3} \int \!\! d^3x \, e^{i {\bf k} 
\cdot ({\bf x} - {\bf x}')} \langle \delta_{\rho}(\eta_s,{\bf x}) 
\delta_{\rho}(\eta_s,{\bf x}') \rangle \approx \frac{\kappa^4 k}{25920\, \pi^4} 
\left(-|k \eta_0|^3 + \frac3{\pi} |k \eta_0|^2 + \cdots \right) \,. 
\label{eq:power_kin}
\end{equation}
Note that if we take the $k^4\, |\eta_0|^3$ term seriously for all $k$,
then its spatial Fourier transform contributes a term proportional
to  $\nabla^4 \delta({\bf x} - {\bf x}')$ in $\langle
 \delta_\rho(\eta_s,{\bf x})\, \delta_\rho(\eta_s,{\bf x'})\rangle$.
To the extent that measurements are made in position space  by
comparing proxies for $\delta \rho(t,{\bf x})$ at 
${\bf x} \neq {\bf x}'$, this term will not contribute. Here we
assume that we may ignore this term and retain only the nonlocal
$\eta_0^2$ effect. Recalling the definition Eq.~(\ref{power}), our result for 
the one loop contribution to the $\delta \rho$ power spectrum from 
conformal matter is then
\begin{equation}
\Bigl[ {\cal P}_{\delta_{\rho}}(k)\Bigr]_{\rm conf} \approx 
\frac{\kappa^4 k^4}{2160 \, \pi^4} \left( \frac{k}{H a(t_0)}\right)^2 \, .
\label{eq:cal_P}
\end{equation}
Apart from numerical factors, this is equivalent to Eq.~(48) in 
Ref.~\cite{WNF07}.
We will see in the next section that the prefactor of $k^4$ should really 
be $H^4$, however, that still leaves the result very strongly biased toward 
short wavelength perturbations and far from scale invariant. The possible
implications will be discussed in Sect.~\ref{sec:density}. 

\section{A Dynamical Model}
\label{sec:dyn}

The computation of $[{\cal P}_{\delta_{\rho}}(k)]_{\rm conf}$ we have just
completed involved three basic steps:
\begin{itemize}
\item{Inferring the post-inflationary density contrast from the
perturbed expansion
\begin{equation}
\delta_{\rho}(t,{\bf x}) = -(1 + w) \int_{t_r}^t \! dt' \, 
\delta \theta(t,{\bf x}) \, . \label{app1}
\end{equation}}
\item{Approximating the post-inflationary Raychaudhuri equation as
(\ref{eq:theta1}), so that the density contrast during radiation domination
becomes
\begin{equation}
\delta_{\rho}(t,{\bf x}) \approx -\frac{2}{3 H(t_r)} \, 
\delta \theta(t_r,{\bf x}) \, . \label{app2}
\end{equation}}
\item{Computing the perturbed expansion which is accumulated during 
inflation by using the Raychaudhuri equation (\ref{Ray}), under the 
assumption that the acceleration term $(u^{\mu}_{~ ;\nu} u^{\nu})_{;\mu}$ 
makes no contribution at linearized order.}
\end{itemize}
Equation~(\ref{app1}) is exact. While  Eq.~(\ref{app2}) is certainly not 
exact, it does represent a reasonable approximation when Fourier transformed
and restricted to super-horizon modes. The problematic step is ignoring the
acceleration term to compute the $\delta \theta(t_r,{\bf x})$ induced by
conformal matter fluctuations during inflation. It turns out that the
acceleration term depends linearly upon the inflaton perturbation, and we
must study the coupled gravity-inflaton system to get a reliable result for
$\delta \theta(t_r,{\bf x})$. Having done this, we use Eqs.~(\ref{app1}) and
(\ref{app2}) as before to compute $[{\cal P}_{\delta_{\rho}}(k)]_{\rm conf}$.

In this section we study the coupling of a single scalar inflaton
field with the gravitational perturbations of a spatially flat
Robertson-Walker spacetime. These perturbations are in turn driven
by the fluctuations of the stress tensor of a conformal quantum field.
The unperturbed metric is of the form in Eq.~(\ref{eq:RW}).
During inflation, this metric will be approximately that of de Sitter
spacetime, although with a slowly varying Hubble parameter $H$.
We assume that the unperturbed metric satisfies Einstein's
equations with the stress tensor of a spatially homogeneous inflaton
field $\varphi_0(t)$ as the source. Let the inflaton be self-coupled 
by a potential $V(\varphi)$. Then the Einstein equations become
\begin{equation}
3 H^2  =  \frac{1}{2} \kappa^2\, \left( \frac{1}{2} \dot{\varphi}_0^2 
+ V_0 \right)\, , \label{eq:Einstein0}
\end{equation}
and 
\begin{equation}
-2 \dot{H} - 3 H^2 = 
\frac{1}{2} \kappa^2\, \left( \frac{1}{2} \dot{\varphi}_0^2 
- V_0 \right)\,. \label{eq:Einstein0'}
\end{equation}
Here dots again denote derivatives with respect to $t$, $H=\dot{a}/a$
and $V_0 =  V(\varphi_0)$. The scalar field equation is
\begin{equation}
-\frac1{\sqrt{-g}} \partial_{\mu} \Bigl( \sqrt{-g} g^{\mu\nu} 
\partial_{\nu} \varphi\Bigr) + V'(\varphi) =0\,, \label{eq:scalar}
\end{equation}
which becomes
\begin{equation}
 \ddot{\varphi}_0 + 3 H \delta \dot{\varphi}_0
 + V'_0 =0\,.   \label{eq:scalar0}
\end{equation}
To the extent that $V_0$ is not constant, the unperturbed spacetime
will not be exactly de Sitter space.

\subsection{Coupled Equations for Inflaton and Metric Perturbations}
\label{sec:coupled}

We next wish to consider linear perturbations of this spacetime
in a gauge in which 
\begin{equation}
g_{tt} =-1\,,
\end{equation}
so the perturbed metric may be written as
\begin{equation}
ds^2 = -dt^2 + 2 a(t) h_{ti}(t,{\bf x}) dt dx^i + a^2(t) \Bigl[\delta_{ij}
+ h_{ij}(t,{\bf x}) \Bigr] dx^i dx^j \; . \label{gmn}
\end{equation}
It is convenient to define the conformally transformed, spatial metric,
\begin{equation}
\widetilde{g}_{ij} \equiv \delta_{ij} + h_{ij} \; .
\end{equation}
The determinant of the full metric can be broken up into three factors,
\begin{equation}
-g = a^6 \times {\rm det}(\widetilde{g}) \times \Bigl[1 + h_{ti} h_{tj} 
\widetilde{g}^{ij} \Bigr] = a^6 \Bigl[1 + h + O(h^2)\Bigr] \; , 
\label{eq:gexpand}
\end{equation}
where $h= \delta^{ij} \, h_{ij}$ is the trace of the metric
perturbation.

In addition to the metric perturbation, the inflaton field will have
inhomogeneous perturbations:
\begin{equation}
\varphi(t,{\bf x}) = \varphi_0(t) + \delta \varphi(t,{\bf x}) \; .
\end{equation}
The scalar field perturbations will satisfy 
\begin{equation}
 \delta \ddot{\varphi} + 3 H \delta \dot{\varphi}
+ \frac12 \dot{\varphi}_0 \dot{h} + V''_0 \delta \varphi  = 0 \,,
\label{eq:scalar1}
\end{equation}
which follows from the expansion of Eq.~(\ref{eq:scalar}) to first
order both in $\delta \varphi$ and in $h$.

The Einstein equations may be written as 
\begin{equation}
R_{\mu\nu} =  \frac{1}{2} \kappa^2\, \left( T^{\rm total}_{\mu\nu} 
- \frac12 g_{\mu\nu}
g^{\rho\sigma} T^{\rm total}_{\rho\sigma}\right) \; ,
\label{eq:Einstein}
\end{equation}
where the total stress tensor is the sum of contributions from the
inflaton field and the conformal quantum field:
\begin{equation}
T^{\rm total}_{\mu\nu} =T^{\rm infl}_{\mu\nu}+ T^{\rm conf}_{\mu\nu}\,. 
\end{equation}
We focus here on the time-time component of the Einstein equation. The
first order expansion of $R_{tt}$ is
\begin{equation}
R_{tt} = -3\dot{H} - 3 H^2 -\frac12 \Bigl(\ddot{h} + 2 H \dot{h}
\Bigr) + \frac1{a} \Bigl(\dot{h}_{ti , i} + H h_{ti , i}\Bigr) + 
O(h^2) \; .
\end{equation}
The inflaton stress tensor satisfies
\begin{equation}
T^{\rm infl}_{\mu\nu} - \frac12 g_{\mu\nu} g^{\rho\sigma} T^{\rm infl}_{\rho
\sigma} = \partial_{\mu} \varphi \partial_{\nu} \varphi + g_{\mu\nu} 
V(\varphi) \; .
\end{equation}
The first order expansion of the $tt$ component of this expression is
\begin{equation}
\partial_{t} \varphi \partial_{t} \varphi - V(\varphi) =
\dot{\varphi}_0^2 - V(\varphi_0) + 2 \dot{\varphi}_0 \delta \dot{\varphi} -
V'(\varphi_0) \delta \varphi  \; .
\end{equation}
Thus the equation for the first order metric perturbation can be
written as 
\begin{equation}
-\frac12 \Bigl(\ddot{h} + 2 H \dot{h} \Bigr) + \frac1{a} \Bigl(
\dot{h}_{ti , i} + H h_{ti , i}\Bigr) = \frac{\kappa^2}{2} \Bigl[2 
\dot{\varphi}_0 \delta \dot{\varphi} - V'(\varphi_0) \delta \varphi + U  
\Bigr] \; , \label{eq:metric1}
\end{equation}
where we define
\begin{equation}
U =  T^{\rm conf}_{tt}\,,
\end{equation}
the energy density of the conformal field in the comoving frame.

Define the normal vector to the surfaces of constant $\varphi$ by
\begin{equation}
u^{\mu} = -\frac{g^{\mu\nu} \partial_{\nu} \varphi}{\sqrt{-g^{\alpha\beta}
\partial_{\alpha} \varphi \partial_{\beta} \varphi}} \; . \label{vel}
\end{equation}
The first order expansion of the spatial components of this vector
is
\begin{equation}
u^i  =  \frac{h_{ti}}{a} + \frac{\partial_i \delta \varphi}{a^2 
\dot{\varphi}_0} \,.
\end{equation}
We can impose the gauge condition
\begin{equation}
u^\mu = \delta^\mu_t\,,  \label{eq:gauge}
\end{equation}
from which the condition $g_{tt}=-1$ follows. In addition, this leads
to a relation between $\delta \varphi$ and $h_{ti}$ to first order:
\begin{equation}
h_{ti}(t,\mathbf{x}) = -\partial_i 
\left[ \frac{\delta \varphi(t,\mathbf{x})}{a(t)
\dot{\varphi}_0(t)} \right]  \; . \label{h0i}
\end{equation}
This relation allows us to eliminate the $h_{ti}$ terms in
  Eq.~(\ref{eq:metric1}) and write
\begin{equation}
-\kappa^2 \dot{\varphi}_0 \delta \dot{\varphi} + \frac12 \kappa^2 V_0'
\delta \varphi - \frac{\nabla^2}{a^2} \frac{\partial}{\partial t} \left(
\frac{\delta \varphi}{\dot{\varphi}_0}\right) - \frac1{2 a^2} 
\frac{\partial}{\partial t} \left( a^2 \dot{h} \right)  =  \frac12 
\kappa^2 \, U \; . \label{eq:metric2}
\end{equation}

Equations (\ref{eq:scalar1}) and (\ref{eq:metric2}) form a pair of
coupled second order equations for the metric and scalar field
perturbations. These equations may be rewritten by expressing 
${\varphi}_0$, $V_0$, and their derivatives in terms of the Hubble
parameter $H(t)$ and its derivatives. The sum of
Eqs.~(\ref{eq:Einstein0}) and (\ref{eq:Einstein0'}) leads to
\begin{equation}
\kappa \dot{\varphi}_0 = 2 \sqrt{-\dot{H}} \; , \label{phidot}
\end{equation}
and their difference leads to
\begin{equation}
\kappa^2 V_0 = 2 \dot{H} + 6 H^2 \; .
\end{equation}
From these relations, we find
\begin{equation}
\kappa \ddot{\varphi}_0 = -\frac{\ddot{H}}{\sqrt{-\dot{H}}} \qquad , \qquad
\kappa \varphi_0^{\hspace{-.25cm} \lower.3ex \hbox{$\cdot\!\!\cdot\!\!\cdot$}}
= -\frac{H^{\hspace{-.25cm} \raise.6ex \hbox{$\cdot\!\!\cdot\!\!\cdot$}}}{
\sqrt{-\dot{H}}} - \frac{\ddot{H}^2}{2 (-\dot{H})^{\frac32}} \; ,
\end{equation}
and 
\begin{equation}
\kappa V_0' = \frac{\ddot{H}}{\sqrt{-\dot{H}}} - 6 H \sqrt{-\dot{H}} 
\qquad , \qquad V_0'' = -\frac{H^{\hspace{-.25cm} \raise.6ex 
\hbox{$\cdot\!\!\cdot\!\! \cdot$}}}{2 \dot{H}} + \frac{\ddot{H}^2}{4 
{\dot{H}}^2} - \frac{3 H \ddot{H}}{2 \dot{H}} - 3 \dot{H} \; .
\end{equation}

Note that the homogeneous form (setting $U =0$) of
Eqs.~(\ref{eq:scalar1}) and (\ref{eq:metric2}) has a solution
when $\delta \varphi = \dot{\varphi}_0$ and $\dot{h} = 6 \dot{H}$.
We can reduce the order of the system by scaling out this
solution and defining
\begin{equation}
B(x) = \frac{\delta \varphi(x)}{\dot{\varphi}_0(t)} \; . \label{new1}
\end{equation}
We can now rewrite  Eq.~(\ref{eq:scalar1}) as
\begin{equation}
\dot{h} = 6 \dot{H} B - 2\left( 2 \frac{\ddot{\varphi}_0}{
\dot{\varphi}_0} + 3 H\right) \dot{B} -2 \ddot{B} \; . \label{eq:h}
\end{equation}
This allows us to eliminate $h$ from Eq.~(\ref{eq:metric2}), and write 
\begin{equation}
{\cal O}\; \dot{B} = \frac12 \kappa^2 \, U \,, 
\label{eq:Phieqn} 
\end{equation}
where ${\cal O}$ is the operator defined by
\begin{equation}
{\cal O} =
\left[ \partial_t^2 + \left(5 H + 2 \frac{\ddot{\varphi}_0}{\dot{\varphi}_0}
\right) \partial_t + 4\dot{H} + 6 H^2 + 4 H \frac{\ddot{\varphi}_0}{
\dot{\varphi}_0} - 2 \frac{\ddot{\varphi}_0^2}{\dot{\varphi}_0^2} +
2 \frac{\varphi_0^{\hspace{-.25cm} \raise.2ex \hbox{$\cdot\!\!\cdot\!\! 
\cdot$}}}{\dot{\varphi}_0} - \frac{\nabla^2}{a^2}\right] \,. \label{eq:opO}
\end{equation}
Equation~(\ref{eq:Phieqn}) is a third-order equation which we will
solve using a retarded Green's function. 

It will be convenient to take a spatial Fourier transform, and 
define the operator
\begin{equation}
{\cal O}_k =
 \partial_t^2 + \left(5 H + 2 \frac{\ddot{\varphi}_0}{\dot{\varphi}_0}
\right) \partial_t + 4\dot{H} + 6 H^2 + 4 H \frac{\ddot{\varphi}_0}{
\dot{\varphi}_0} - 2 \frac{\ddot{\varphi}_0^2}{\dot{\varphi}_0^2} +
2 \frac{\varphi_0^{\hspace{-.25cm} \raise.2ex \hbox{$\cdot\!\!\cdot\!\! 
\cdot$}}}{\dot{\varphi}_0} + \frac{k^2}{a^2} \,. \label{eq:opOk}
\end{equation} 
Let $G(t,t',k)$ be the retarded Green's function of this operator,
which satisfies the equation
\begin{equation}
{\cal O}_k\, G(t,t',k) = \delta(t-t') \,,
\end{equation}
with the boundary condition
\begin{equation}
 G(t,t',k) = 0 \qquad {\rm if} \qquad t <t'  \,.
\end{equation}
Let $\Psi_1$ and $\Psi_2$ be two linearly independent solutions of
the homogeneous equation
\begin{equation}
{\cal O}_k\, \Psi(t,k) = 0 \,.  \label{eq:homog}
\end{equation}
The Green's function may be expressed as
\begin{equation}
G(t,t',k) = \frac{1}{W(t',k)} \left[
\Psi_1(t_<,k) \Psi_2(t_>,k) - \Psi_1(t,k) \Psi_2(t',k)\right] \; ,
\end{equation}
where $t_<$ and $t_>$ are the lesser and the greater, respectively, of $t$
and $t'$, and 
\begin{equation}
W(t,k) = \Psi_1(t,k) \dot{\Psi}_2(t,k) - \dot{\Psi}_1(t,k) \Psi_2(t,k) 
\end{equation}
is the Wronskian.

The homogeneous solutions $\Psi_i$ are difficult to obtain in general.
However, if we make a ``slow roll'' approximation in which time
derivatives of $H$, and hence of $\varphi_0$ and of $V_0$, are
assumed to be small, then we have approximately
\begin{equation}
{\cal O}_k \approx
 \partial_t^2 + 5 H \, \partial_t + 6 H^2 + \frac{k^2}{a^2} \,. \label{eq:Ok_SR}
\end{equation} 
In this approximation, the solutions of Eq.~(\ref{eq:homog}) are
\begin{equation}
\Psi_1(t,k) = a^{-2}(t)\, {\rm e}^{ik \int_{t_0}^t dt_1 a^{-1}(t_1)} \,,
\end{equation}
and  $\Psi_2(t,k) = \Psi^*_1(t,k)$. Here $t_0$ is an arbitrary
constant. Now the Wronskian is
\begin{equation}
W(t,k) = -\frac{2ik}{a^5(t)}\,, 
\end{equation} 
and the retarded Green's function may be written as
\begin{equation}
G(t,t',k) = -\frac{a^3(t')}{k\, a^2(t)}\; 
\sin\left[k \int_{t'}^t \frac{dt_1}{a(t_1)} \right] \,, \label{eq:G_SR}
\end{equation}
for $t \geq t'$.

\subsection{Inflaton Field Fluctuations}

In this subsection, we wish to calculate the fluctuations in $\varphi$
which are driven by the stress tensor fluctuations of the conformal
field. Let $G(x,x')$ be a coordinate space Green's function for the
operator $\cal{O}$, which satisfies
\begin{equation}
{\cal O}\, G(x,x') = \delta(x-x') = 
\delta(t-t')\,\delta({\bf x} -{\bf x}')\,.
\end{equation} 
A particular solution of Eq.~(\ref{eq:Phieqn}) can be written as 
\begin{equation}
\dot{B}(t,{\bf x}) = \frac12 \kappa^2 \int d^4x_1\,  G(x,x_1) \,U(x_1)\,.
\end{equation}

We now treat $U$ and hence $B$ as fluctuating fields, and write the
correlation function for $\dot{B}$ as
\begin{equation}
\langle \dot{B}(t,{\bf x}) \dot{B}(t,{\bf x}') \rangle =
\frac14 \kappa^4 \int d^4x_1 d^4x_2\,  G(x,x_1) G(x',x_2) \,
\langle U(x_1) U(x_2) \rangle \,.
\end{equation}
Next we convert from comoving to conformal time, using $d\eta =
dt/a(t)$, and use the relation between the (comoving) energy density in
Robertson-Walker spacetime to that in flat spacetime,
\begin{equation}
\langle U(x_1) U(x_2) \rangle = a^{-4}(\eta_1)\, a^{-4}(\eta_2)\,
{\cal E}(\Delta \eta,r)\,,
\end{equation}
where ${\cal E}(\Delta \eta,r)$ is the flat spacetime energy density
correlation function, with $r=|{\bf x}_1 -{\bf x}_2|$. This leads to
\begin{eqnarray}
\langle \partial_\eta B(\eta,{\bf x}) \,
\partial_\eta B(\eta',{\bf x}') \rangle &=&
\frac14 \kappa^4 a(\eta) a(\eta') 
\int_{\eta_0}^\eta \frac{d\eta_1}{a^3(\eta_1)}  
 \int_{\eta_0}^{\eta'} \frac{d\eta_2}{a^3(\eta_2)} \int d^3x_1
d^3x_2\,       \nonumber \\
& \times & \left[ G(\eta,\eta_1,{\bf x} -{\bf x}_1)\,
 G(\eta',\eta_2,{\bf x}' -{\bf x}_2)\, {\cal E}(\Delta \eta,r)\right]\,.
\end{eqnarray}
Here the boundary condition $\partial_\eta B(\eta,{\bf x})=0$ at
$\eta = \eta_0$ has been imposed.

Next we take spatial Fourier transforms, and define
\begin{equation}
\langle \partial_\eta B(\eta,{\bf x}) \,
\partial_{\eta'} B(\eta',{\bf x}') \rangle = 
\int d^3k\; {\rm e}^{i{\bf k}\cdot ({\bf x} -{\bf x}')}
\, \langle \partial_\eta B \,\partial_{\eta'} B \rangle_k \,,
\end{equation}
and analogous transforms of $ G(\eta,\eta_1,{\bf x} -{\bf x}_1)$
and of ${\cal E}(\Delta \eta,r)$. Then we may write
\begin{eqnarray}
\langle \partial_\eta B \,\partial_{\eta'} B \rangle_k &=&
\frac{1}{4} \kappa^4 a(\eta) a(\eta')
\int_{\eta_0}^\eta \frac{d\eta_1}{a^3(\eta_1)}  
 \int_{\eta_0}^{\eta'} \frac{d\eta_2}{a^3(\eta_2)}        \nonumber \\
&\times& \left[ G(\eta,\eta_1,k)\,
 G(\eta',\eta_2,k)\, \hat{\cal{E}}(\Delta \eta,k)\right]\,.
\end{eqnarray} 
The correlation function for $\partial_\eta B$ may be integrated
to yield the mean squared inflaton fluctuation at the end of
inflation, $\eta=\eta_r$. If  $B=0$ at $\eta = \eta_0$, then
\begin{equation}
\langle B^2(\eta_r) \rangle_k = \int_{\eta_0}^{\eta_r} d\eta
\int_{\eta_0}^{\eta_r} d\eta' 
\langle \partial_\eta B\,\partial_{\eta'} B \rangle_k \,.
\end{equation}
 
The form for the Green's function from the slow roll approximation,
Eq.~(\ref{eq:G_SR}), may be expressed as
\begin{equation}
G(\eta,\eta',k) = -\frac{a^3(\eta')}{k\, a^2(\eta)}\; 
\sin[k(\eta -\eta')]  \,. \label{eq:G_SR2}
\end{equation}
With this form, and the de Sitter space scale factor, 
Eq.~(\ref{eq:deSitscale})
we find
\begin{eqnarray}
\langle B^2(\eta_r) \rangle_k &=& \frac{\kappa^4 H^2}{4 k^2}
\int_{\eta_0}^{\eta_r} d\eta\, \eta \int_{\eta_0}^{\eta_r} d\eta' \,\eta'
\int_{\eta_0}^{\eta} d\eta_1\, \sin[k(\eta -\eta_1)]   \nonumber \\
&\times& \int_{\eta_0}^{\eta'} d\eta_2\, \sin[k(\eta' -\eta_2)]\;
 \hat{\cal {E}}(\Delta \eta,k) \,. \label{eq:Phisq}
\end{eqnarray}
It is interesting to compare this with the result 
Eq.~(\ref{eq:theta_corr}) of 
the kinematical model. After first horizon crossing, $\delta \hat{\theta}(
\eta,{\bf k}) \sim H^2 \hat{B}(\eta,{\bf k})$ [See Eq.~(\ref{eq:dth})
below.], so the big difference between 
Eqs.~(\ref{eq:Phisq}) and (\ref{eq:theta_corr}) is the extra factors of
\begin{equation}
\frac{H^2}{k} \int_{\eta_0}^{\eta} \!\! d\eta_1 \, \sin[k(\eta \!-\! \eta_1)]
\times \frac{H^2}{k} \int_{\eta_0}^{\eta'} \!\! d\eta_2 \,
\sin[k(\eta' \!-\! \eta_2)] \; .
\end{equation}
These terms describe how stress tensor fluctuations from very early times
are communicated by the inflaton field to the late time geometry, and
they effectively introduce
a factor of $(H/k)^2 \times (H/k)^2 = (H/k)^4$ to the result of the
kinematical model.

Finally, we may use the form of $\hat{\cal {E}}(\Delta \eta,k)$ given
in Eq.~(\ref{eq:em_corr_k}), and perform the integrations in 
Eq.~(\ref{eq:Phisq})
using the algebraic computer program {\it Mathematica}. The result
is rather complicated, but in the limit of large $|\eta_0|$, it
becomes
\begin{equation}
\langle B^2(\eta_r) \rangle_k \sim 
-\frac{\kappa^4\, H^2\, |\eta_0|^3}{122880 \pi^4} +
\frac{\kappa^4\, H^2\, \eta_0^2}{153600 \pi^3\, k} + ... \;. 
\label{eq:asymp}
\end{equation}
As with our result Eq.~(\ref{eq:F0_asym}) for $\hat{F}_0(k)$, the leading
contribution for large $|\eta_0|$ corresponds to a term which is 
ultralocal in position space, in this case proportional to 
$\delta({\bf x} - {\bf x}')$.

In the subsequent analysis, we will also encounter expectation values
of quadratic forms involving time derivatives of $B$, such as
$\langle B \dot{B} \rangle$ and $\langle  \dot{B}^2 \rangle$.
However, one may check that all of these terms are at most 
proportional to  $|\eta_0|$ and hence sub-dominant
compared to $\langle B^2 \rangle$ in the limit of large $|\eta_0|$.

\subsubsection{Density fluctuations from the conservation law}

As discussed at the beginning of this section, we can compute the 
density contrast using Eqs.~(\ref{app1}) and (\ref{app2}) once we have
$\delta \theta(t_r,{\bf x})$, the perturbed expansion at the end of 
inflation. The expansion $\theta$ can be obtained from its definition 
\begin{equation}
\theta \equiv u^{\mu}_{~ ;\mu} = \frac1{\sqrt{-g}} \partial_{\mu} \Bigl(
\sqrt{-g} \, u^{\mu} \Bigr) \; . \label{def}
\end{equation}
In our gauge, Eq.~(\ref{eq:gauge}), this becomes
\begin{equation}
\theta(t,{\bf x}) = \frac{\partial}{\partial t} \ln\Bigl(\sqrt{-g} \, 
\Bigr) \; .
\end{equation}
We may use Eq.~(\ref{eq:gexpand}) to write
\begin{equation}
\theta = 3 H + \frac12 \frac{\partial}{\partial t} \ln(\widetilde{g}) +
\frac12 \frac{\partial}{\partial t} \ln\Bigl[1 + h_{ti} h_{tj} 
\widetilde{g}^{ij} \Bigr] = 3 H + \frac12 \dot{h}      + O(h^2) \; .
\end{equation}
Recall that $\theta_0 = 3H$ is the expansion of the comoving
geodesics in Robertson-Walker spacetime, so the first order
perturbation of the expansion is 
\begin{equation}
\delta \theta(t,{\bf x}) = \frac12 \dot{h}(t,{\bf x}) \; .
\label{eq:theta-h}
\end{equation}

One infers $\dot{h}(t,{\bf x})$ from $B(t,{\bf x})$ using
Eq.~(\ref{eq:h}). Because we only need it at the end of inflation the
$\dot{B}$ and $\ddot{B}$ terms can be dropped, and we can use the 
radiation domination result $\dot{H} = -2 H^2$ to conclude
\begin{equation}
\delta \theta(t_r,{\bf x}) \approx -6 H^2 B(t_r,{\bf x}) \, . \label{eq:dth}
\end{equation}
Hence Eqs.~(\ref{app1}) and (\ref{app2}) give the following expression
for the density contrast during radiation domination:
\begin{equation}
\delta_{\rho}(t,{\bf x}) \approx -4 H B(t_r,{\bf x}) \, . \label{eq:dcon}
\end{equation}
This should be valid when Fourier transformed and restricted to 
super-horizon modes.

To find the power spectrum we first compute the spatial Fourier transform 
of the $\delta_{\rho}$ correlator using Eqs.~(\ref{eq:dcon}) and
(\ref{eq:asymp})
\begin{equation}
P_{\delta \rho}(k) \equiv \frac1{(2\pi)^3} \int \!\! d^3x \, e^{i {\bf k} 
\cdot ({\bf x} - {\bf x}')} \langle \delta_{\rho}(\eta_s,{\bf x}) 
\delta_{\rho}(\eta_s,{\bf x}') \rangle \approx \frac{\kappa^4 H^4 k^{-3}}{
7680\, \pi^4} \left(-|k \eta_0|^3 + \frac{4\pi}{5} |k \eta_0|^2 + \cdots\right)
\,. \label{eq:power_dyn}
\end{equation}
Multiplying by $4 \pi k^3$ gives the power spectrum. As for the
kinematic model (\ref{eq:power_kin}), we assume that we may drop the 
$|\eta_0|^3$ term which is ultralocal and presumably not part of the
observed power spectrum. This leaves us with
\begin{equation}
\Bigl[ {\cal P}_{\delta_{\rho}}(k)\Bigr]_{\rm conf} \approx 
\frac{\kappa^4 H^4}{2400 \, \pi^2} \left( \frac{k}{H a(t_0)}\right)^2 \, .
\label{eq:final}
\end{equation}
Thus the dynamical model also produces a non-scale invariant spectrum 
biased toward the blue end of the spectrum, although less so than in the 
case of the kinematic model.

\subsubsection{Density fluctuations from the Sachs-Wolfe effect}
\label{sec:SW}

An alternative approach to calculate density or temperature
fluctuations is to study the effects of metric perturbations on
the redshifts of photons, as was first done by Sachs and Wolfe~\cite{SW67}.
Equation~(39) of their paper may be expressed as
\begin{equation}
\frac{\Delta T}{T} = \int_{t_r}^{t_s} dt \left[\hat{e}^i h_{ti,t}(x)
-\frac{1}{2} \hat{e}^i \hat{e}^j\, h_{ij,t}(x) \right]\,.
\label{eq:SW}
\end{equation}
This formula gives the differential redshift, and hence temperature
fluctuation, of a photon propagating from $t=t_r$ to $t=t_s$ along
a null geodesic in the direction of the unit vector $\hat{e}^i$. The
integrand is understood to be evaluated along the unperturbed null
geodesic. In contrast to the previous discussion, we now need
expressions for the individual components of the spatial metric
perturbation, $h_{ij}$. For this purpose, it is convenient to express
the scalar part of the perturbed metric, Eq.~(\ref{gmn}) as
\begin{equation}
ds^2 = -dt^2 -2 B_{,i} dx^i dt +\left[a^2(1-2\psi) \delta_{ij}
-2 E_{,ij} \right] dx^i dx^j\,. \label{eq:gmn2}
\end{equation}
Here we follow the notation of Mukhanov~\cite{Mukhanov}, as
modified in Ref.~\cite{W09}.
As before, $h_{ti}$ is given by Eq.~(\ref{h0i}). The spatial
components of scalar perturbations are
\begin{equation}
h_{ij}(t,\mathbf{x}) = -2 \delta_{ij}\, \psi(t,\mathbf{x}) 
-\frac{2}{a^2(t)}\, \partial_i \partial_j\, E(t,\mathbf{x})\,. 
\end{equation}
The quantities which appear in the integrand of Eq.~(\ref{eq:SW})
may be written in terms of $B$, $\psi$ and $E$ as
\begin{equation}
\hat{e}^i h_{ti,t}(x) = 
-\left(\frac{\hat{\mathbf{e}}\cdot \mathbf{\nabla}}{a}\right)
(\dot{B}-HB)
\label{eq:SW1}
\end{equation}
and
\begin{equation}
-\frac{1}{2} \hat{e}^i \hat{e}^j\, h_{ij,t}(x) = \dot{\psi}
+\left(\frac{\hat{\mathbf{e}}\cdot \mathbf{\nabla}}{a}\right)^2 (\dot{E}-2HE)\,.
\end{equation}

Expressions for $\psi$ and for $\dot{E}-2HE$, Eqs.~(\ref{eq:psi}) 
and (\ref{eq:E}),
respectively, are derived in the Appendix. First we note that
$\dot{E}-2HE$ contains two types of terms, those which involve 
$\dot{B}$ and $\ddot{B}$, and those which depend upon $U$ or $\dot{U}$
evaluated at the same time as $\dot{E}-2HE$. Both of these types of
terms will give a sub-dominant contribution, which is either
independent of $|\eta_0|$ or small compared to the $|\eta_0|^3$ and
$|\eta_0|^2$ terms. The same comment applies to all terms in $\psi$, 
except for the $H\,B$ term. Note that in coordinate space,  
Eqs.~(\ref{eq:psi}) and (\ref{eq:E}) contain the non-local operator 
$1/\nabla^2$, which is non-local in space only, not in time. In any case,
calculations are best done in Fourier space, where $1/\nabla^2$ is
replaced by $-1/k^2$.

Thus we may take
\begin{equation}
\psi \approx  H\, B\,.
\end{equation}
If we drop the $\dot{B}$ term in Eq.~(\ref{eq:SW1}), and use the fact
that
\begin{equation}
\frac{d}{dt} = \frac{\partial}{\partial t} - 
\left(\frac{\hat{\mathbf{e}}\cdot \mathbf{\nabla}}{a}\right)
\end{equation}
is the total derivative along our null geodesic, we may write
\begin{equation}
\frac{\Delta T}{T} = \int_{t_r}^{t_s} dt \; \frac{d\, (HB)}{dt}
\approx -(HB)_{t_r}\,.
\end{equation}
In the last step, we used the fact that the dominant contribution
will come from the lower limit of the integral. If we recall that
here 
\begin{equation}
\frac{\Delta \rho}{\rho} = 4\;\frac{\Delta T}{T}\,,
\end{equation}
we again obtain Eq.~(\ref{eq:dcon}).

Density perturbations are often treated using the gauge invariant
potentials, which is yet another possible approach. However,
both the fluid flow approach and the Sachs-Wolfe formula,
Eq.~(\ref{eq:SW}), are themselves gauge invariant and somewhat simpler
for our purposes than the gauge invariant potentials.

\section{Density Perturbations from Quantum Stress Tensor Fluctuations}
\label{sec:density}

\subsection{Possible Constraints on the Duration of Inflation}

Let us now discuss the possible physical implications of the
conformal matter contribution to the power spectrum. Recalling that 
$\kappa^2 = 16 \pi G$, and that current wave numbers $k_{\rm now}$ 
correspond to $k = (a k)_{\rm now}$, our result (\ref{eq:final}) can 
be expressed as
\begin{equation}
\Bigl[ {\cal P}_{\delta_{\rho}}\Bigr]_{\rm conf} \approx \frac{8 G^2 H^4}{75}
\left(\frac{(a k)_{\rm now}}{a_0 H}\right)^2 \, . \label{eq:fin}
\end{equation}
This is not scale invariant, and it is associated with highly non-Gaussian
fluctuations. In contrast, observations of large scale structure and the 
cosmic microwave background radiation are consistent with the primordial 
perturbation spectrum being approximately scale invariant and Gaussian 
\cite{WMAP}
\begin{equation}
{\cal P}_{\mathcal{R}}(k_{\rm now}) \approx \Bigl(2.441^{+0.088}_{-0.092}\Bigr)
\times 10^{-9} \left(\frac{k_{\rm now}}{0.002~{\rm Mpc}^{-1}}\right)^{
-0.037 \pm 0.012} \, . \label{eq:obs}
\end{equation}
(The primordial curvature and density contrast power spectra are 
related by ${\cal P}_{\mathcal{R}} = \frac9{16} {\cal P}_{\delta_{\rho}}$
\cite{Mukhanov}.) Note that the weak scale dependence of the observed
power spectrum (\ref{eq:obs}) is actually in the opposite (red) sense to the 
massive blue tilt we predict from conformal matter. Hence the contribution
from conformal matter can only represent a tiny part of the total power
spectrum. Because our result Eq.~(\ref{eq:fin}) grows like $1/a^2(t_0)$ as the
start of inflation is pushed back to earlier and earlier times, one can 
derive a bound on the duration of inflation by requiring that 
Eq.~(\ref{eq:fin}) is small enough to not affect the measured result,
 Eq.~(\ref{eq:obs}).

It will facilitate the discussion to recall some reasonably generic 
predictions of single-scalar inflation in the slow roll approximation.
The tree order results for the scalar and tensor power spectra are \cite{LL93}
\begin{equation}
\Bigl[{\cal P}_{\mathcal{R}}(k_{\rm now})\Bigr]_{\rm tree} \approx
\frac{G H^2(t_k)}{\pi \epsilon(t_k)} \qquad , \qquad
\Bigl[{\cal P}_{h}(k_{\rm now})\Bigr]_{\rm tree} \approx \frac{16}{\pi} \,
G H^2(t_k) \, , \label{eq:theory}
\end{equation}
where $\epsilon(t) \equiv -\dot{H}/H^2$, and $t_k$ is the time of first 
horizon crossing,
\begin{equation}
a_{\rm now} k_{\rm now} = a(t_k) H(t_k) \, .
\end{equation}
The absence of much scale dependence in the observed 
result, Eq.~(\ref{eq:obs}),
is explained by $H(t)$ being approximately constant during inflation. (This
is why our de Sitter approximation of Sect.~\ref{sec:dyn} was well motivated.)
Of course a nearly constant $H(t)$ makes the slow roll parameter $\epsilon(t)
= -\dot{H}/H^2$ close to zero. The enhancement of the scalar power spectrum
by $1/\epsilon(t_k)$ explains why it has been observed, while the tensor
contribution has so far not been resolved. At 95\% confidence the bound on
their ratio is \cite{WMAP}
\begin{equation}
r \equiv \frac{{\cal P}_{h}(0.002~{\rm Mpc}^{-1})}{{\cal P}_{\mathcal{R}}(
0.002~{\rm Mpc}^{-1})} < 0.22 \, .
\end{equation}
With the theoretical results Eq.~(\ref{eq:theory}) and the scalar observation
Eq.~(\ref{eq:obs}), this implies a bound on the inflationary Hubble parameter
\begin{equation}
G H^2 = \frac{\pi}{16} \times r \times {\cal P}_{\mathcal{R}}(
0.002~{\rm Mpc}^{-1}) \alt 10^{-10} \, .
\end{equation}
Note that our one loop contribution Eq.~(\ref{eq:fin}) is suppressed
by $G H^2\epsilon$ relative to the tree effect Eq.~(\ref{eq:theory}). 
It can only become
observable when inflation begins at such an early time that these factors
are canceled by the square of the physical wave number in Hubble units,
$(k/a_0 H)^2$.

The bound we get on $t_0$ derives from requiring the predicted contribution
from conformal matter Eq.~(\ref{eq:fin}) to be smaller than the
observed result Eq.~(\ref{eq:obs}) for the smallest wave number 
$k_{\rm now}$ for which data 
exists. We take $k_{\rm now} \approx 10^{-24} {\rm cm^{-1}} \approx 2 \times 
10^{-38}~{\rm GeV}$, which corresponds to structures of about $2~{\rm Mpc}$
in physical size, or about $5$ arcminutes of angular scale~\cite{ACBAR}.
Let $T_R$ stand for the reheat temperature, and let us assume efficient 
reheating so that
\begin{equation}
H^2 \approx 8 \pi G T_R^4 \approx 8\pi \times 10^{10}~{\rm GeV}^2 
\left(\frac{T_R}{10^{12}~{\rm GeV}}\right)^4 \, .
\end{equation}
The Universe has expanded by about a factor of $10^3$ since the last 
scattering time $t_s$ (when the temperature was $T_S \approx 1~{\rm eV}$) 
and recall that we normalize the scale factor to unity at the end of 
inflation, hence
\begin{equation}
a_{\rm now} \approx 10^3 \,a(t_s) \approx 10^3\, \frac{T_R}{T_S} \approx
10^{24} \left(\frac{T_R}{10^{12}~{\rm GeV}}\right) \,.
\end{equation}
For these parameters our result Eq.~(\ref{eq:fin}) implies
\begin{equation}
\Bigl[ {\cal P}_{\mathcal{R}}(k_{\rm now})\Bigr]_{\rm conf} \approx 
\frac{3 G^2 H^4}{50} 
\left(\frac{a_{\rm now} \, k_{\rm now}}{a_0 \, H}\right)^2 \approx
\frac{5 \times 10^{-94}}{a_0^2} 
\left(\frac{T_R}{10^{12}~{\rm GeV}}\right)^6 \, .
\end{equation}
Requiring this conformal contribution to be less (by a factor of ten, say) 
than the observed spectrum Eq.~(\ref{eq:obs}) gives
\begin{equation}
\frac1{a_0} \alt 10^{42} \left(\frac{10^{12}~{\rm GeV}}{T_R}\right)^3 \, .
\label{eq:bound}
\end{equation}
Recall that sufficient inflation to solve the horizon and flatness problems
requires $1/a_0 \agt 10^{23}$, so Eq.~(\ref{eq:bound}) allows more than 
enough inflation for this purpose. Note that this bound is, apart from
being improved by a factor of $10^3$, equivalent to Eq.~(92) in 
Ref.~\cite{WNF07}. However, the latter result was derived using an
overly simplified dynamical model which did not fully account for the
coupling between the inflaton field and the perturbations of the
spacetime geometry.

\subsection{The Transplanckian Issue}

We now turn to some of the conceptual issues which are raised by the
calculations described in the previous sections. One of these concerns
the use of transplanckian modes, that is, modes whose physical wavelengths 
are less than the Planck length of $\ell_p \equiv \sqrt{G} \approx 
10^{-33}~{\rm cm}$, as measured by an observer at the start of inflation. 
The estimates in the previous subsection dealt with perturbations with a
present wavelength on the order of $\lambda_{\rm now} \approx 10^{25}\,
{\rm cm} \approx 10^{58} \ell_p$. The physical wavelength of this mode at
the beginning of inflation is,
\begin{equation}
\lambda_0  = a_0 \lambda = a_0 \times \frac{\lambda_{\rm now}}{a_{\rm now}} 
\, .
\end{equation}
If we assume that $a_0$ is at the bound (\ref{eq:bound}) --- which means
conformal matter contributes $10\%$ of the measured power spectrum at the
smallest observed scales --- then the initial wavelength is
\begin{equation}
\lambda_0 \approx 10^{-8} \, \ell_p\, 
\left(\frac{T_R}{10^{12}~{\rm GeV}}\right)^2 \,.
\end{equation}
For all reasonable values of $T_R$, this is far below the Planck length.  

This raises two questions:
\begin{itemize}
\item{Is it valid to extrapolate low energy dynamics such as
electromagnetism to transplanckian scales?}
\item{Is it valid to apply perturbation theory for transplanckian modes?}
\end{itemize}
No one knows what dynamical principles might apply at Planck scales, but it
is of course acceptable to carry out a study, as we have done, based on the
explicitly stated assumption that they are unchanged. What doesn't seem
alright is employing perturbation theory. One must not be mislead by the
fact that the tree order effect $\sim \kappa^2 H^2/\epsilon$ is small; the
series is an expansion in powers of the large parameter $(\kappa k/a_0)^2$,
\begin{equation}
{\cal P}_{\delta_{\rho}}(k) \sim \kappa^2 H^2 \Biggl\{\frac{\alpha_0}{\epsilon} 
+ \alpha_1 \Bigl(\frac{\kappa k}{a_0}\Bigr)^2 + 
\alpha_2 \Bigl(\frac{\kappa k}{a_0}\Bigr)^4 + \dots\Biggr\} \; . \label{series}
\end{equation}
If one makes the usual assumption that the pure numbers $\alpha_{\ell}$ are
of order one then the only way of making the one loop term comparable to
the tree order result must also make the two loop and higher terms
comparable. The conclusion seems unavoidable that perturbation theory
must break down, for mode $k$, as the initial time is pushed back to the
point for which $\kappa k/a_0 \sim 1$. We do not possess a nonperturbative
computational technique so what actually happens at earlier times is a
matter of conjecture and lively debate within the community \cite{BM,KN,many}.

One view is based on the observation that the far ultraviolet contains so
many modes that even very small deviations from quiescence in each of them
must produce enormous fluctuations that would invalidate semi-classical
inflation. Hence it must be, the argument goes, that a nonperturbative
resummation of loop corrections such as Eq.~(\ref{series}) exhibits no large
effect, even for very early initial times. For each wave number $k$ there
would be a time $T_k$ such that $\kappa k/a(T_k) \ll 1$, after which our
perturbative treatment is valid. As long as $t_0$ comes after $T_k$, making
$t_0$ smaller causes the one loop effect to grow as we predict, with higher
loop contributions still negligble. But if $t_0$ is pushed before $T_k$
then the higher loop corrections become important and the whole series
approaches a constant. If this view is correct then, for $t_0 < T_k$, one
could only employ the perturbative treatment of this paper by starting the
evolution of $\hat{B}(t,{\bf k})$ at $t = T_k$, not at $t = t_0$. And the
correct initial condition would be something close to quiescence at $t =
T_k$. This is a nonlocal initial condition, but then quantum effects
typically are nonlocal.

A different view is motivated by the similarity of these issues to those 
which arise in black hole physics. The original derivation of the Hawking 
effect~\cite{Hawking} assumes free quantum field theory on a fixed 
background spacetime and requires transplanckian modes. This derivation 
is analogous to our treatment in the previous sections. It is possible to 
reproduce the Hawking effect without the use of transplanckian 
modes~\cite{Unruh,CJ96}, but only by postulating a non-linear dispersion 
relation, which breaks local Lorentz symmetry. 

If there is a new physical principle which avoids transplackian modes,
then economy of thought would suggest that it should be the same
principle for both black hole physics and for cosmology. Ideally,
one might hope for an experimental or observational test of
transplanckian physics. The power spectrum which we have derived
using transplanckian modes has the potential to provide such a
test. If inflation lasted just slightly less than the amount
given by Eq.~(\ref{eq:bound}), then the model described above 
predicts a non-scale invariant and non-Gaussian component in the cosmic 
microwave background which might be detectable.

\subsection{Relation to Weinberg's Theorem}

Neither the comoving wave number $k$ nor scale factor $a(t)$ are
physical, only their ratio, $k/a(t)$. Of course ratios of the scale
factor at different times are also physical. Because we normalize
the scale factor to one at the end of inflation, the various factors
of $k$ in our results must really be interpreted as the physical wave
number at the end of inflation, $k/a_r$. Therefore, the kinematic model
estimate of $[{\cal P}_{\delta_{\rho}}(k)]_{\rm conf} \sim 
\kappa^4 k^6/[H^2 a_0^2]$ seemes to suggest a one loop correction to 
the power spectrum which not only violates scale invariance by the 
factor $(k/a_r)^6$ but also grows at late times like $(a_r/a_0)^2$. 
Such growth would contradict a bound of at most logarithmic growth 
established by Weinberg~\cite{Weinberg}. (Weinberg's result was derived 
for minimally coupled scalars but it can easily be extended to conformally 
coupled particles~\cite{Chai07}.) In fact one can see from the dynamical 
model of Sect.~\ref{sec:dyn} that there is no growth at late times; what
happens instead is that the principal effect arises from fluctuations 
near the time $t_0$ when the interaction is turned on, after which it 
rapidly approaches a constant. By comparing our one loop correction with 
the usual tree order result
\begin{equation}
\Bigl[{\cal P}_{\delta_{\rho}}(k)\Bigr]_{\rm tree} \sim 
\frac{\kappa^2 H^2}{\epsilon} \qquad {\rm versus} \qquad 
\Bigl[{\cal P}_{\delta_{\rho}}(k)\Bigr]_{\rm conf} \sim \kappa^2 H^2 \times
\frac{\kappa^2 k^2}{a_0^2} \; ,
\end{equation}
It will be seen that our contribution consists of the tree result
(without the inverse of $\epsilon \equiv -\dot{H}/H^2$), multiplied by
a typical one loop correction of the square of $\kappa$ times the mode's
physical energy at the initial time. Later times contribute
far less because the mode's physical energy redshifts so rapidly.
There is no mystery about why the effect can be large at very early
times because the mode is transplanckian then and should induce large
gravitational effects. Of course this again raises concerns about using
perturbation theory and low energy dynamics. What Weinberg considered was
quantum loop effects from the ``safe'' regime of late times during which
perturbative general relativity must be valid. Our results are in perfect
agreement with his bound; indeed, they fail to show even the logarithmic
growth allowed for by the bound and achieved by nonconformal matter.

\section{Summary and Discussion}
\label{sec:final}

In this paper we have evaluated the extra contribution to inflationary
density perturbations from the quantum stress tensor fluctuations of a
conformal field such as the photon. This was done in a simple,
kinematical model and then in a more accurate, but much more complicated,
dynamical model. Our main result is that the power spectrum of the
energy density at wave number $k$ goes like the tree order result
(without enhancement by $1/\epsilon$) times $(E(t_0)/M_{\rm Pl})^2$,
where $E(t_0) = k/a(t_0)$ is the mode's physical energy at the beginning
of inflation and $M_{\rm Pl}$ is the Planck mass. If a perturbative
computation such as this could be trusted to arbitrarily early times,
the absence of such a massive blue tilt in the observed power spectrum
would seem to imply a bound on the total duration of inflation. This
constraint allows enough inflation to solve the horizon and flatness
problems.

Our result derives from very early times and rapidly approaches a constant,
so it does not contradict Weinberg's bound~\cite{Weinberg,Chai07}, on the 
maximum possible growth of quantum corrections at late times. However, our 
result does involve a problematic extrapolation of known physical laws to 
the transplanckian regime, and the even more problematic assumption that
perturbation theory can be used at times and on modes for which the
physical energy density is transplanckian. Opinion on these issues is
divided~\cite{BM,KN,many} and we have tried to present both sides. It is 
worth pointing out that stress tensor fluctuations from very early times 
would also induce significant non-Gaussianities if one were to compute 
them perturbatively, using known physical laws, as we have done for the 
2-point correlator.

\begin{acknowledgments}
We have benefited from discussions with many colleagues. L.H.F.
would especially like to thank the participants of the 14th Peyresq
workshop for lively discussions. This work was supported in part by 
FONDECYT grant number 3100041, by National Science Foundation Grants
PHY-0653085, PHY-0855021 and PHY-0855360, by the Institute for
Fundamental Theory at the University of Florida, and by the National 
Science Council, Taiwan, ROC under the Grant NSC 98-2112-M-001-009-MY3 (KWN).

The Centro de Estudios Cient\'ificos (CECS) is funded by the Chilean
Government through Millennium Science Initiative, the Centers of
Excellence Base Financing Program of Conicyt and Conicyt grant 
"Southern Theoretical Physics Laboratory" ACT-91. CECS is also supported
by a group of private companies which at present includes Antofagasta
Minerals, Arauco, Empresas CMPC, Indura, Naviera Ultragas, and
Telef\'onica del Sur. CIN is funded by Conicyt and the Gobierno Regional
de Lo R\'ios.
\end{acknowledgments}

\appendix

\section{Some relations involving spatial perturbations}

In this appendix, we will derive some relations relating to the
spatial parts of the metric perturbations which are used in 
Sect.~\ref{sec:SW}. The Einstein equations, Eq.~(\ref{eq:Einstein}),
may be expressed as
\begin{equation}
R_{\mu\nu} - \frac12 \kappa^2 \Bigl[\partial_{\mu} \varphi \partial_{\nu}
\varphi + g_{\mu\nu} V(\varphi)\Bigr] = \frac12 \kappa^2 
T_{\mu\nu}^{\rm conf} \; . \label{eq:EE}
\end{equation}
It is convenient to remove the scale factors from the conformal 
stress tensor by defining
\begin{equation}
U = T_{tt}^{\rm conf} = \hat{T}_{tt} \quad , \quad T_{ti}^{\rm conf}
= a \hat{T}_{ti} \quad , \quad T_{ij}^{\rm conf} = a^2 
\hat{T}_{ij} \; .
\end{equation}
Note that $\hat{T}_{\mu\nu}$ are the components of the conformal
stress
tensor in a local orthonormal frame defined by $d\hat{t} =dt$ and
$d\hat{x^i} = a\, dx^i$. The $h_{ti}$ and $h_{ij}$ defined in
Eq.~(\ref{gmn}) are the metric perturbations in this frame.

Because the conformal stress tensor is itself a first order perturbation
we can express its tracelessness using only the zeroth order metric,
\begin{equation}
g^{\rho\sigma} T_{\rho\sigma}^{\rm conf} = 0 \quad \Longrightarrow \quad
\hat{T}_{kk} = \hat{T}_{tt} + O(h^2) \; . \label{trrel}
\end{equation}
Similar simplifications can be made to the relations implied by 
stress-energy conservation,
\begin{eqnarray}
g^{\rho\sigma} T_{\mu \rho ; \sigma}^{\rm conf} = 0 \qquad 
& \Longrightarrow & \nonumber \\
\frac1{a^3} \partial_t (a^3 \hat{T}_{tt}) & = & \frac1{a} 
\hat{T}_{tk ,k} - H \hat{T}_{kk} + O(h^2) \; , \\
\frac1{a^3} \partial_t (a^4 \hat{T}_{ti}) & = & \hat{T}_{i k , k}
+ O(h^2) \; . \label{SE2}
\end{eqnarray}

The expansion of the time-time component of Eq.~(\ref{eq:EE}) was
performed in Sect.~\ref{sec:coupled} so here we focus on the 
remaining components.
The first order expansions of the required components of the Ricci tensor 
are, using the metric of Eq.~(\ref{gmn}),
\begin{eqnarray}
\lefteqn{R_{ti} = \dot{h}_{k [i , k]} + \frac1{a} h_{t [k , i] k}
+ (3 H^2 + \dot{H}) a h_{ti} + O(h^2) \; , } \\
\lefteqn{R_{ij} = (3 H^2 + \dot{H}) a^2 \delta_{ij} + (3 H^2 + \dot{H})
a^2 h_{ij} + \frac1{2 a} \partial_t (a^3 \dot{h}_{ij})
- \frac1{a} \partial_t (a^2 h_{t (i , j)} ) } \nonumber \\
& & + h_{k (i , j) k} - \frac12 h_{ij , kk} - \frac12 h_{,i j}
- H a \delta_{ij} h_{t k , k} + \frac12 H a^2 \delta_{ij} \dot{h}
+O(h^2) \; . \qquad 
\end{eqnarray}
The remaining first-order Einstein equations become
\begin{equation}
\dot{h}_{k [i , k]} + 2 \dot{H} \partial_i \Phi = \frac12 \kappa^2 a
\hat{T}_{ti} \; , \label{1st}
\end{equation}
and
\begin{eqnarray}
\lefteqn{\Bigl[\partial_t^2 + 3 H \partial_t - \frac{\nabla^2}{a^2}\Bigr] 
h_{ij} + \frac1{a^2} \Bigl[ h_{ik , kj} + h_{jk , ki} - h_{,ij}\Bigr] 
+ H \delta_{ij} \dot{h} } \nonumber \\
& & -2 \delta_{ij} \Bigl[\ddot{H} + 6 H \dot{H}\Bigr] B
+ \frac2{a^2} \dot{B}_{, ij} + \frac{2 H}{a^2} \Bigl[\delta_{ij} 
\nabla^2 + \partial_i \partial_j\Bigr] B = \kappa^2 \hat{T}_{ij} 
\; . \qquad \label{2nd}
\end{eqnarray}

To understand what Eqs.~(\ref{1st}) and (\ref{2nd}) imply, it is
useful to make a decomposition of $h_{ij}$ 
into irreducible representations of the rotation group,
\begin{equation}
h_{ij} \equiv h^{TT}_{ij} + h^T_{i , j} + h^T_{j , i} - \frac12 \Bigl(
\delta_{ij} - 3 \frac{\partial_i \partial_j}{\nabla^2} \Bigr) h^L
+ \frac12 \Bigl( \delta_{ij} - \frac{\partial_i \partial_j}{\nabla^2}
\Bigr) h \; .
\end{equation}
This is a decomposition into transverse-tracefree (TT), transverse
(T), longitudinal (L) and trace parts.
Here  $h^{TT}_{ii} = 0 = h^{TT}_{ij , j}$ and $h^T_{i , i} = 0$ as usual.
We can make similar decompositions of the conformal stress tensor,
\begin{eqnarray}
\hat{T}_{ti} & \equiv & T^{T}_{ti} + \partial_i T^{L}_t \; , \\
\hat{T}_{ij} & \equiv & T^{TT}_{ij} + T^T_{i , j} + T^T_{j , i} - 
\frac12 \Bigl(\delta_{ij} - 3 \frac{\partial_i \partial_j}{\nabla^2} \Bigr) 
T^L + \frac12 \Bigl( \delta_{ij} - \frac{\partial_i \partial_j}{\nabla^2}
\Bigr) T \; . \qquad
\end{eqnarray}
The following identities facilitate extraction of the longitudinal and 
trace parts,
\begin{eqnarray}
\delta_{ij} & = & -\frac12 \Bigl(\delta_{ij} - 3 \frac{\partial_i 
\partial_j}{\nabla^2}\Bigr) + \frac32 \Bigl(\delta_{ij} - \frac{\partial_i
\partial_j}{\nabla^2}\Bigr) \; , \\
\partial_i \partial_j & = & -\frac12 \Bigl(\delta_{ij} - 3 \frac{\partial_i 
\partial_j}{\nabla^2}\Bigr) \nabla^2 + \frac12 \Bigl(\delta_{ij} - 
\frac{\partial_i \partial_j}{\nabla^2}\Bigr) \nabla^2 \; .
\end{eqnarray}
Equivalently, the longitudinal part is obtained by the action of the
projection operator
\begin{equation}
L_{ij} = \frac{\partial_i\,\partial_j}{\nabla^2}\,,
\end{equation}
so that $h^L = L_{ij}\, h_{ij}$.
The longitudinal part of (\ref{2nd}) is,
\begin{eqnarray}
\lefteqn{\Bigl( \partial_t^2 + 3 H \partial_t + \frac{\nabla^2}{a^2} \Bigr)
h^L + \Bigl(H \partial_t - \frac{\nabla^2}{a^2}\Bigr) h} \nonumber \\
& & \hspace{2cm} -\Bigl(2\ddot{H} + 12 H \dot{H}\Bigr) B + 
\frac{\nabla^2}{a^2} \Bigl(2\partial_t + 4 H \Bigr) B= 
\kappa^2 T^L \; . \qquad \label{long}
\end{eqnarray}

Note that stress-energy conservation (\ref{SE2}) implies,
\begin{equation}
\Bigl(\partial_t + 3 H\Bigr) (a T^{L}_t)  = T^L \; , \label{hL0}
\end{equation}
and Eq.~(\ref{1st}) implies 
\begin{equation}
\dot{h}^L  =  \dot{h} - 4 \dot{H} \, B + \kappa^2 a T^{L}_t \; .
\label{0i2}
\end{equation}
\begin{eqnarray}
\lefteqn{\frac{\nabla^2}{a^2} h^L + \Bigl(\partial_t^2 + 4 H \partial_t - 
\frac{\nabla^2}{a^2}\Bigr) h} \nonumber \\
& & \hspace{1cm} -4 \dot{H} \dot{B} -\Bigl(6\ddot{H} + 24 H \dot{H}\Bigr) 
B + \frac{\nabla^2}{a^2} \Bigl(2\partial_t + 4 H \Bigr) B = 
0 \; . \qquad \label{newlong}
\end{eqnarray}
Now eliminate the $\dot{h}$ terms using Eq.~(\ref{eq:h}) and 
eliminate the resulting $\partial_t^3 B$ term using  Eq.~(\ref{eq:Phieqn}).
The resulting simplification of (\ref{newlong}) is,
\begin{equation}
\frac{\nabla^2}{a^2} \Bigl(h^L - h + 4 H B\Bigr) + \Bigl(-8 H 
\frac{\ddot{\varphi}_0}{\dot{\varphi}_0} - 12 H^2 + 4 \dot{H}\Bigr)
\dot{B} - 4 H \ddot{B} = \kappa^2 \,U \; . \label{eq:long2}
\end{equation}

At this point, it is convenient to switch to the variables $\psi$ and
$E$ defined in Eq.~(\ref{eq:gmn2}), in terms of which
\begin{equation}
h^L = -2\psi -2 \frac{\nabla^2}{a^2} E\,, 
\end{equation}
and
\begin{equation}
h = -6\psi -2 \frac{\nabla^2}{a^2} E\,. \label{eq:h_psi}
\end{equation}
Now Eq.~(\ref{eq:long2}) may be written as
\begin{equation}
\psi - HB + \frac{a^2}{\nabla^2} \left[H \ddot{B} +
\left(2H \frac{\ddot{\varphi}_0}{\dot{\varphi}_0} +3H^2
  -\dot{H}\right)\dot{B} +\frac{\kappa^2}{4} U \right]=0\,. \label{eq:psi}
\end{equation}
The time derivative of Eq.~(\ref{eq:h_psi}) is
\begin{equation}
\dot{h} = -6\dot{\psi} -2 \frac{\nabla^2}{a^2} (\dot{E}-2HE)\,.
\end{equation}
Next substitute this relation and the time derivative of
Eq.~(\ref{eq:psi}) into Eq.~(\ref{eq:h}), and eliminate the
$\partial_t^3 B$ term using  Eq.~(\ref{eq:Phieqn}). The result is
\begin{equation}
\frac{\nabla^2}{a^2} (\dot{E}-2HE) = \ddot{B} + 
\left(\frac{\ddot{H}}{\dot{H}} +3H\right) \dot{B}
- \kappa^2 \frac{a^2}{\nabla^2}\left(3HU +\frac{3}{4}\dot{U}
\right)\,.
\label{eq:E}
\end{equation}


\begin{thebibliography}{99}

\bibitem{MC81} V. Mukhanov and G. Chibisov, JETP Lett. {\bf 33}, 532 (1981).

\bibitem{GP82} A.H. Guth and S.-Y. Pi, Phys. Rev. Lett. {\bf 49}, 1110 (1982).

\bibitem{Hawking82} S.W. Hawking, Phys. Lett. B {\bf 115}, 295 (1982).

\bibitem{Starobinsky82}  A.A. Starobinsky,
Phys. Lett. B {\bf 117}, 175 (1982).

\bibitem{BST83}  J.M. Bardeen, P.J. Steinhardt, and
M.S. Turner, Phys. Rev. D {\bf 28}, 679 (1983).

\bibitem{Mukhanov} V. Mukhanov, {\it Physical Foundations of
Cosmology}, (Cambridge University Press, 2005).

\bibitem{WMAP} E. Komatsu, {\it et al.}, Astrophys. J. Suppl, {\bf
180}, 330 (2009), arXiv:0803.0547; arXiv:1001.4538.

\bibitem{WF01} C.-H. Wu and L.H. Ford, Phys. Rev. D {\bf 64}, 045010 (2001),
quant-ph/0012144.

\bibitem{Borgman} J. Borgman and L.H. Ford, Phys. Rev. D {\bf 70}
 064032 (2004), gr-qc/0307043.

\bibitem{Stochastic} B.L. Hu and E. Verdaguer, Living Rev. Rel. {\bf 7},
3 (2004), gr-qc/0307032.


\bibitem{FW04} L.H. Ford and R.P. Woodard, Class. Quant. Grav. {\bf 22}, 
1637 (2005), gr-qc/0411003.

\bibitem{TF06} R.T. Thompson and L.H. Ford,  Phys. Rev. D {\bf 74},
024012 (2006), gr-qc/0601137.

\bibitem{FW07} L.H. Ford and C.H. Wu, AIP Conf.Proc. {\bf 977} 145
  (2008), arXiv:0710.3787.

\bibitem{FFR10}  C.J. Fewster, L.H. Ford, and T.A. Roman,
  arXiv:1004.0179.



\bibitem{WNF07} C.H. Wu, K.W. Ng, and L.H. Ford,  Phys. Rev. D 
{\bf 75}, 103502 (2007), arXiv:gr-qc/0608002. 

\bibitem{BV94} A. Borde and A. Vilenkin, Phys. Rev. Lett.
{\bf 72}, 3305 (1994). 

\bibitem{BGV03} A. Borde, A.H. Guth, and A. Vilenkin, Phys. Rev. Lett.
{\bf 90}, 151301 (2003), gr-qc/0110012. 

\bibitem{W10} S. Winitzki, arXiv:1003.1680.

\bibitem{Hawking66} S.W. Hawking, Astrophys. J. {\bf 145}, 544 (1966).

\bibitem{Olson} D.W. Olson,  Phys. Rev. D {\bf 14}, 327 (1976).

\bibitem{LS90} D.H. Lyth and E.D. Steward, Astrophys. J. {\bf 361}, 343 (1990).

\bibitem{LL93} A.D. Liddle and D.H. Lyth, Phys. Rep. {\bf 231}, 1 (1993),
astro-ph/9303019.

\bibitem{ACBAR} C.L. Kuo {\it et al}, Astrophys. J. {\bf 600}, 32 (2004),
astro-ph/0212289.

\bibitem{SW67} R.K. Sachs and A.M. Wolfe, Astrophys. J. {\bf 147}, 73 (1967).

\bibitem{W09} R. P. Woodard, Rept. Prog. Phys. {\bf 72}, 126002 (2009),
arXiv:0907.4238.

\bibitem{BM} R. H. Brandenberger and J. Martin, Mod. Phys. Lett. {\bf A16}
(2001) 999, astro-ph/0005432;
Phys. Rev. {\bf D63} (2001) 123501, hep-th/0005209;
Phys. Rev. {\bf D66} (2002) 083514, hep-th/0112122;
Phys. Rev. {\bf D65} (2002) 103514, hep-th/0201189;
Int. J. Mod. Phys. {\bf A17} (2002) 3663, hep-th/0202142;
Phys. Rev. {\bf D68} (2003) 063513, hep-th/0305161;
Phys. Rev. {\bf D71} (2005) 023504, hep-th/0410223.

\bibitem{KN} J. C. Niemeyer, Phys. Rev. {\bf D63} (2001) 123502,
astro-ph/0005533;
A. Kempf, Phys. Rev. {\bf D63} (2001) 083514, astro-ph/0009209;
A. Kempf and J. C. Niemeyer, Phys. Rev. {\bf D64} (2001) 103501,
astro-ph/0103225.

\bibitem{many} J. C. Niemeyer and R. Parentani, Phys. Rev. {\bf D64}
(2001) 101301, astro-ph/0101451;
A. A. Starobinsky, JETP Lett. {\bf 73} (2001) 371, astro-ph/0104043;
R. Easther, B. R. Greene, W. H. Kinney and G. Shiu, Phys. Rev. {\bf D64}
(2001) 103502, hep-th/0104102;
Phys. Rev. {\bf D67} (2003) 063508, hep-th/0110226;
Phys. Rev. {\bf D66} (2002) 023518, hep-th/0204129;
L. Hui and W. H. Kinney, Phys. Rev. {\bf D65} (2002) 103507, astro-ph/0109107;
M. Lemoine, Musongela Lubo, J. Martin and J. P. Uzan, Phys. Rev. {\bf D65}
(2002) 023510, hep-th/0109128;
N. Kaloper, M. Kleban, A. E. Lawrence and S. Shenker, Phys. Rev. {\bf D66}
(2002) 123510, hep-th/0201158;
F. Lizzi, G. Mangano, G. Miele and M. Peloso, JHEP {\bf 0206} (2002) 049,
hep-th/0203099;
R. H. Brandenberger and P. M. Ho, Phys. Rev. {\bf D66} (2002) 023517,
hep-th/0203119;
U. H. Danielsson, Phys. Rev. {\bf D66} (2002) 023511, hep-th/0203198;
Phys. Rev. {\bf D71} (2005) 023516, hep-th/0411172;
N. Kaloper, M. Kleban, A. E. Lawrence, S. Shenker and L. Susskind, JHEP
{\bf 0211} (2002) 037, hep-th/0209231;
S. Shankaranarayanan, Class. Quant. Grav. {\bf 20} (2003) 75, gr-qc/0203060;
S. F. Hassan and M. S. Sloth, Nucl. Phys. {\bf B674} (2003) 434,
hep-th/0204110;
J. C. Niemeyer, R. Parentani and D. Campo, Phys. Rev. {\bf D66} (2002)
083510, hep-th/0206149;
K. Goldstein and D. A. Lowe, Phys. Rev. {\bf D67} (2003) 063502,
hep-th/0208167;
C. P. Burgess, J. M. Cline, F. Lemieux and R. Holman, JHEP {\bf 0302}
(2003) 048, hep-th/0210233;
L. Bergstrom and U. H. Danielsson, JHEP {\bf 0212} (2002) 038, hep-th/0211006;
G. L. Alberghi, R. Casadio and A. Trononi, Phys. Lett. {\bf B579} (2004) 1,
gr-qc/0303035;
R. H. Brandenberger, Lect. Notes Phys. {\bf 646} (2004) 127, hep-th/0306071;
J. Martin and C. Ringeval, Phys. Rev. {\bf D69} (2004) 083515,
astro-ph/0310382;
JCAP {\bf 0608} (2006) 009, astro-ph/0605367;
C. P. Burgess, Living Rev. Rel. {\bf 7} (2004) 5, gr-qc/0311082;
Class. Quant. Grav. {\bf 24} (2007) S795, arXiv:0708.2865;
M. S. Sloth, Nucl. Phys. {\bf B748} (2006) 149, astro-ph/0604488;
c. Armedariz-Picon, M. Fontanini, R. Penco and M. Trodden, Class. Quant.
Grav. {\bf 26} (2009) 185002, arXiv:0805.0114.

\bibitem{Hawking} S.W. Hawking, Commun. Math. Phys. {\bf 43}, 199 (1975).

\bibitem{Unruh} W.G. Unruh,  Phys. Rev. D {\bf 51}, 2827 (1994), gr-qc/9409008.

\bibitem{CJ96} S. Corley and  T. Jacobson,  Phys. Rev. D {\bf 54}, 1568 (1996),
hep-th/9601073.

\bibitem{Weinberg} S. Weinberg, Phys. Rev. D {\bf 72}, 043514 (2005);
{\bf 74}, 023508 (2006).

\bibitem{Chai07} K. Chaicherdsukal, Phys. Rev. D {\bf 75}, 063522 (2007).

\end{thebibliography}
\end{document}